\documentclass[a4paper,11pt]{article}
\usepackage{pos}
\usepackage{slashed}

\title{Exploring dark matter models with global fits}

\author*[a]{Tom\'as E. Gonzalo}

\affiliation[a]{Institute for Theoretical Particle Physics (TTP), Karlsruhe Institute of Technology (KIT), 76128 Karlsruhe, Germany}

\emailAdd{tomas.gonzalo@kit.edu}

\abstract{In this conference paper I present the results from a few global studies of Dark Matter (DM) models, in light of recent constraints from direct detection, indirect detection and collider experiments. I show the most recent analysis of models of singlet Higgs-portal DM, where the DM particle is a scalar, vector, Majorana or Dirac fermion. I also present the results from a global study of an effective field theory of DM, where we find that the model shows a strong preference for a low scale of new physics. For all models I show the prospects for detection or exclusions with future experiments.}

\FullConference{%
  8th Symposium on Prospects in the Physics of Discrete Symmetries\\
  7-11 November, 2022\\
  Baden-Baden, Germany
}

\newcommand{\Q}[2]{
  \if\relax\detokenize{#2}\relax
    \mathcal{Q}_{#1}
  \else
    \mathcal{Q}_{#1}^{(#2)}
  \fi
}

\newcommand{\C}[2]{
  \if\relax\detokenize{#2}\relax
    \mathcal{C}_{#1}
  \else
    \mathcal{C}_{#1}^{(#2)}
  \fi
}
\newcommand{\La}{{\rm \Lambda}}

\begin{document}
\maketitle

\section{Introduction}

The evidence for the existence of Dark Matter (DM) is overwhelming. Among the most popular candidates for dark matter are Weakly Interacting Massive Particles (WIMPS), which have the right properties to produce enough abundance to match cosmological observations, and to be detected through interactions with the Standard Model (SM) in laboratory experiments. In spite of the effort of the experimental community, no definite signal has been observed so far, and that has led to strong bounds on the validity of WIMP models. The rigorous combination of existing bounds and predictions for future experiments is therefore a critical endeavour as could determine the survability of WIMPs as DM candidates.

There is a multitude of models that follow the WIMP paradigm, most of them designed to avoid experimental constraints. Some WIMP models are UltraViolet (UV) complete, i.e. they are valid to arbitrary high energies. One such type of models are known as Higgs-portal models, where the only interaction of the DM particle with the SM is through the Higgs~\cite{Athron:2018hpc}. An alternative approach, which allows for the study of multiple models simultaneously, is the Effective Field Theory (EFT) approach. In this case new interactions are parametrised as effective higher dimensional operators between the SM and DM particles~\cite{GAMBIT:2021rlp}. This comes at a price, though, as such models are only valid up to a given energy scale, where the EFT prescription breaks down. An intermediate approach, known as the simplified model approach, parametrises the interactions of DM particles and the SM via the exchange of a mediator particle~\cite{Chang:2022jgo}.

Irrespective of the particular DM model, a rigorous analysis of its validity requires the combination of multiple constraints, the exploration of potentially large and complicated parameter spaces, and a proper statistical interpretation of results. These can all be done within the context of global fits, which combine constraints into a composite likelihood function, explore parameter spaces using smart sampling strategies and allow the interpretation of results in both Bayesian and frequentist frameworks~\cite{AbdusSalam:2020rdj}. The most modern tool that can perform these studies is the Global and Modular Beyond-the-Standard-Model Inference Tool (GAMBIT)~\cite{Athron:2017ard}.

In this conference article I will discuss the results from a global fit, using the GAMBIT software, of a selection of singlet Higgs-portal DM models in Section \ref{sec:HP}, where I would focus on the frequentist results (for the Bayesian results see \cite{Athron:2018hpc}); and an EFT model of DM in Section \ref{sec:DMEFT}.

\section{Higgs-portal singlet DM}
\label{sec:HP}

The various Higgs-portal DM models are differentiated by the DM particle spin, which determines the type of interaction with the Higgs and which constraints are applicable. In all cases the DM particle is a SM singlet, i.e. it does not interact with the gauge bosons of the SM. The Lagrangian of a singlet scalar DM, with a $\mathcal{Z}_2$ symmetry to ensure stability, can be written as
\begin{equation}
 \mathcal{L}_S = \frac{1}{2}\mu_S^2S^2 + \frac{1}{2}\lambda_{hS}S^2|H|^2 + \frac{1}{4}\lambda_S S^4 + \frac{1}{2}\partial_\mu S \partial^\mu S,
\end{equation}
where $S$ is the DM particle, $H$ the Higgs, $\lambda_{hS}$ the dimensionless portal coupling, and $\lambda_S$ and $\mu_S$ the mass parameter and self-coupling of $S$. After electroweak symmetry breaking (EWSB) the mass of $S$ is given by $m_S^2 = \mu_S^2 + \tfrac{1}{2}\lambda_{hS}v_0^2$, with $v_0$ the Higgs vacuum expectation value.

The Lagrangian for the vector DM case is very similar and has the form
\begin{equation}
 \mathcal{L}_V = -\frac{1}{4}W_{\mu\nu}W^{\mu\nu} + \frac{1}{2}\mu_V^2V_\mu V^\mu - \frac{1}{4!}\lambda_V(V_\mu V^\mu)^2+\frac{1}{2}\lambda_{hV}V_\mu V^\mu H^\dagger H,
\end{equation}
where $V$ is the DM particle, $W$ the field strength tensor for $V$, $\lambda_{hV}$ the dimensionless portal coupling, and $\mu_V$ and $\lambda_V$ the mass parameter and self-coupling of $V$. The mass of the vector DM particle $V$ is given by $m_V^2 = \mu_V^2 + \tfrac{1}{2}\lambda_{hV}v_0^2$.

Fermionic DM models are conceptually different from the bosonic ones above, as there are no renormalisable interactions between the Higgs and the DM particle. However, one can write non-renormalisable operators, by introducing a scale $\Lambda$ up to which the effective theory is valid. These models are therefore not purely UV-complete, as they involve effective interactions and a cut-off scale, but they can still be classified as Higgs-portal models as the only, non-renormalisable, interaction of the DM particle is with the Higgs. The Lagrangians for the fermionic DM models are
\begin{align}
 \mathcal{L}_\psi &= \bar\psi(i\slashed\partial - \mu_\psi)\psi - \frac{\lambda_{h\psi}}{\Lambda_\psi}(\cos\theta\bar\psi\psi+\sin\theta\bar\psi i\gamma_5\psi)H^\dagger H, \\
 \mathcal{L}_\chi &= \frac{1}{2}\bar\chi(i\slashed\partial - \mu_\chi)\chi - \frac{1}{2}\frac{\lambda_{h\chi}}{\Lambda_\chi}(\cos\theta\bar\chi\chi + \sin\theta\bar\chi i\gamma_5\chi)H^\dagger H,
\end{align}
where $\psi$ is a Dirac fermion, $\chi$ a Majorana fermion, $\mu_\psi$ and $\mu_\chi$ the mass parameters of the $\psi$ and $\chi$, respectively, $\lambda_{h\psi}$ and $\lambda_{h\chi}$ the portal couplings, $\Lambda_\psi$ and $\Lambda_\chi$ the cut-off scales, and $\theta$ a CP-violating phase. The masses of the fermionic DM candidates are
\begin{equation}
 m_{\chi,\psi} = \left[\left(\mu_{\chi,\psi}+\frac{1}{2}\frac{\lambda_{h\chi,h\psi}}{\Lambda_{\chi,\psi}}v_0^2\cos\theta\right)^2+\left(\frac{1}{2}\frac{\lambda_{h\chi,h\psi}}{\Lambda_{\chi,\psi}}v_0^2\sin\theta\right)^2\right]^{1/2}.
\end{equation}

For each of these models one can make predictions that can be tested by experimental constraints on the model parameters. One of the most relevant constraints is that of their thermal relic abundance. If the DM particles do not annihilate efficiently before they freeze out, their abundance today may be in tension with the measured abundance of $\Omega_{\rm DM}h^2 = 0.1188\pm0.0010$, from the 2015 Planck results~\cite{Planck:2015fie}, the one available at the time this work was performed. For each model and combination of paramters we compute the relic density of DM using \textsf{DarkSUSY}~\cite{Bringmann:2022vra}. As a conservative approach, we do not require our DM candidate to saturate the relic abundance and hence for each DM candidate $X=S,V,\psi,\chi$, we define the DM fraction as $f_X = \Omega_X /\Omega_{\rm DM}$. We use this fraction to scale the spin-independent cross section for direct detection, with a factor of $f_X$, and the annihilation cross section for indirect detection, with a factor of $f_X^2$.

Direct detection of DM as it interacts with matter in underground laboratories is one of the major strategies to search for DM. Using extremely purified materials, it is possible to measure the recoil of a nucleus from the scattering of DM particles. This has historically imposed very strong bounds on the spin-independent cross section for DM. We include direct detection constraints from the experiments LUX 2016, PandaX 2016, CDMSLite, CRESST-II, CRESST-III, PICO-60 and DarkSide-50~\cite{LUX:2016ggv, PandaX-II:2016vec, SuperCDMS:2015eex, CRESST:2015txj, PICO:2017tgi, DarkSide:2018kuk} for the scalar singlet model, and also XENON1T and PandaX 2017~\cite{PandaX-II:2017hlx,XENON:2018voc} for the vector and fermion models. We used \textsf{DDCalc}~\cite{Workgroup:2017lvb} to calculate the likelihood contributions.

Efficient annihilation of DM particles in dense media, such as dwarf spheroidal galaxies, can produce a burst of high-energy particles that are detectable by terrestrial telescopes. The Fermi-LAT sattelite has measured the $\gamma$-rays coming from a combination of 15 dwarf galaxies, \textsf{Pass-8}~\cite{Fermi-LAT:2015att}, which posed strong constraints on the annihilation rate of DM particles. We included this constraint in our analysis using the tool \textsf{gamLike}~\cite{Workgroup:2017lvb}. Captured DM in the Sun can also efficiently annihilate and produce high energy final state neutrinos that can be measured by netrino detectors on Earth. We used \textsf{Capt'n General}~\cite{Kozar:2021iur} to compute the capture rate of DM in the Sun, the produced neutrino flux with \textsf{nulike}~\cite{Scott:2012mq} and compare it with the 79-string IceCube search~\cite{IceCube:2012ugg}.

As these DM candidates interact with the Higgs boson, they can be produced via the decays of Higgs bosons at colliders, if kinematically allowed, i.e. $m_h > 2m_X$. These decays would contribute to the invisible decay width of the Higgs, which is contrained to be less than 19\% at 2$\sigma$ CL~\cite{Belanger:2013xza}.

The final bounds on Higgs-portal models are theoretical constraints. To avoid perturbative unitarity issues in the vector DM model, we take the conservative bound $0 \leq \lambda_{hV} \leq 2m_V^2 / v_0^2$. Fermionic DM models include non-renormalisable operators as portal couplings, and therefore are restricted by the validity of the EFT. Hence, to ensure the perturbativity of the couplings and EFT validity, we assume the constraint $\lambda_{hX}/\Lambda_X \geq 4\pi / 2m_X$, with $X=\chi,\psi$.

\begin{figure}[ht]
\includegraphics[width=0.31\textwidth]{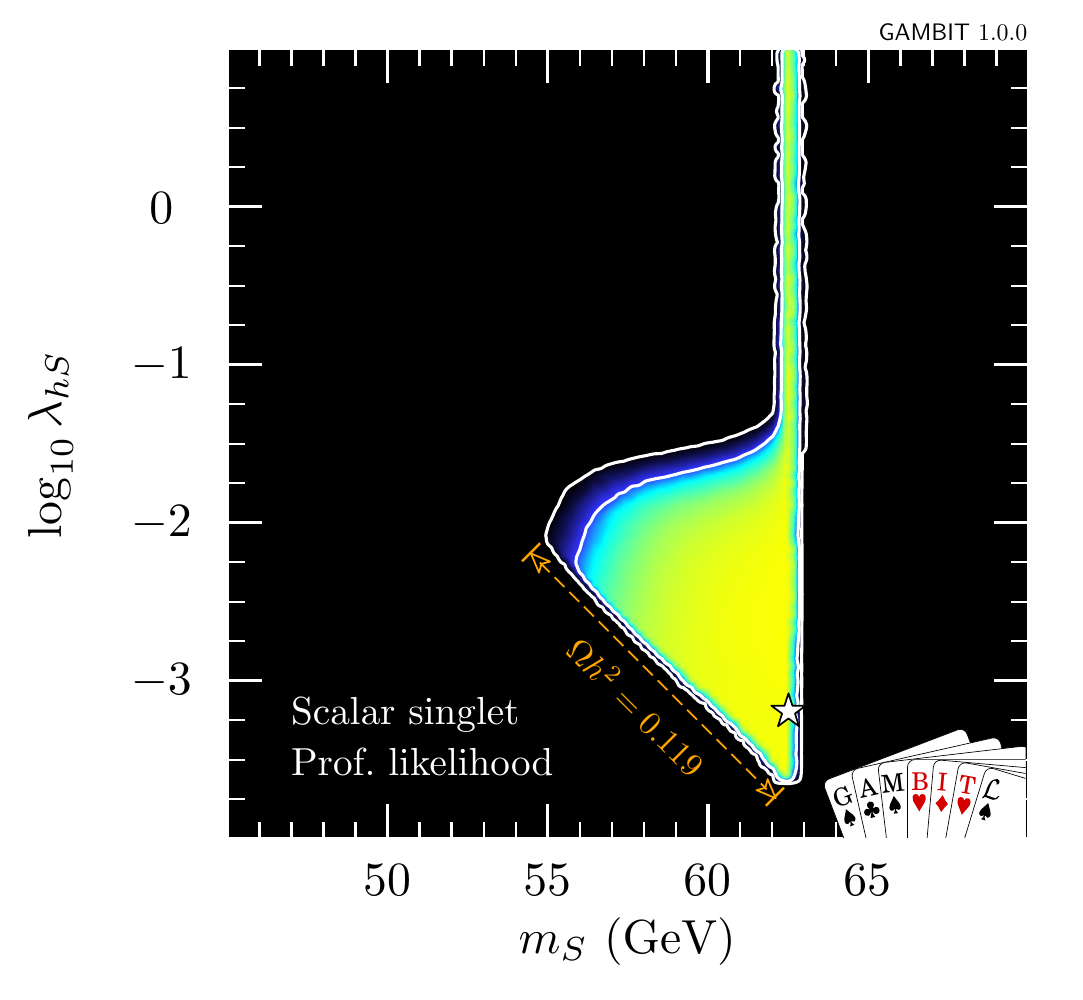}
\includegraphics[width=0.35\textwidth]{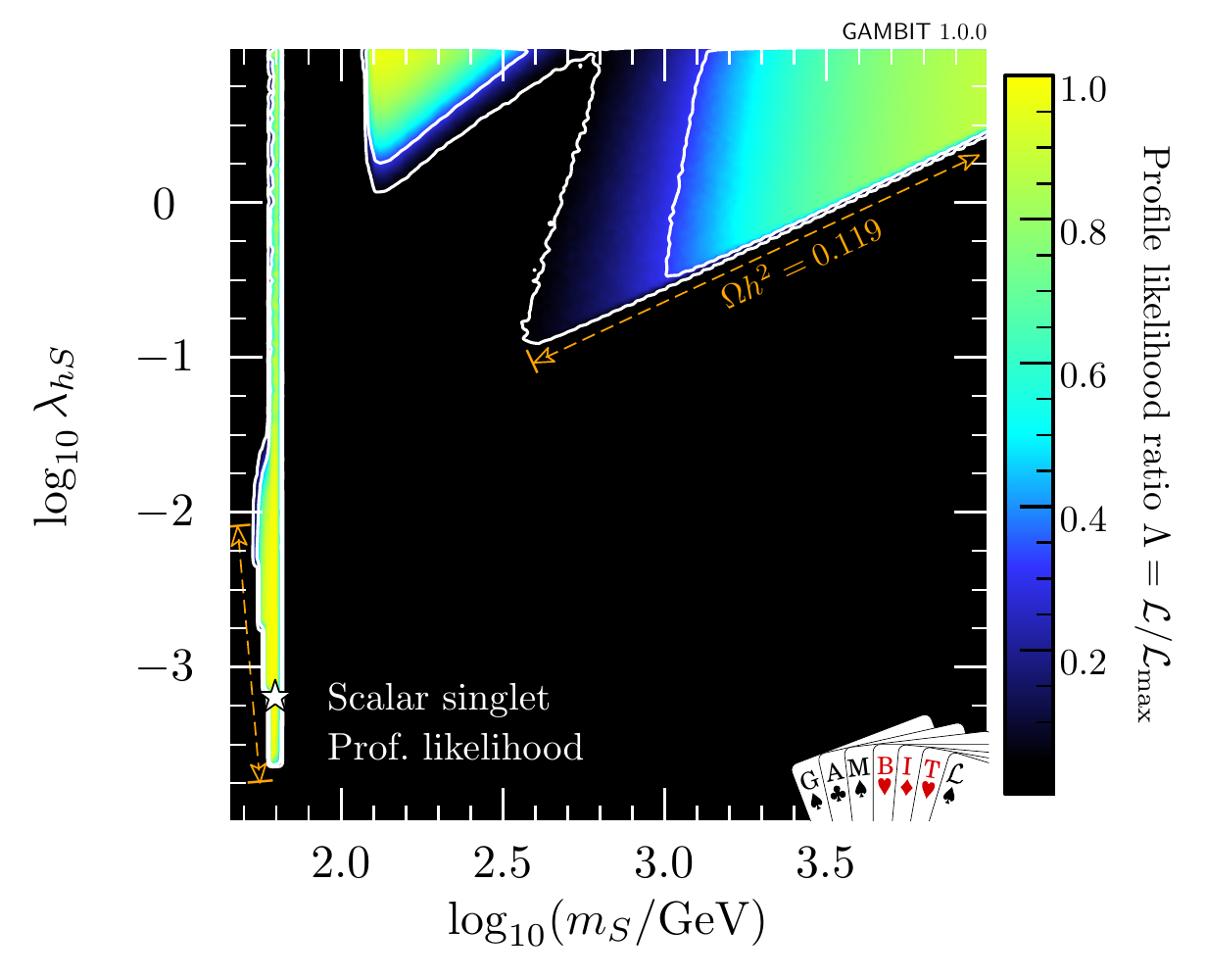}
\includegraphics[width=0.31\textwidth]{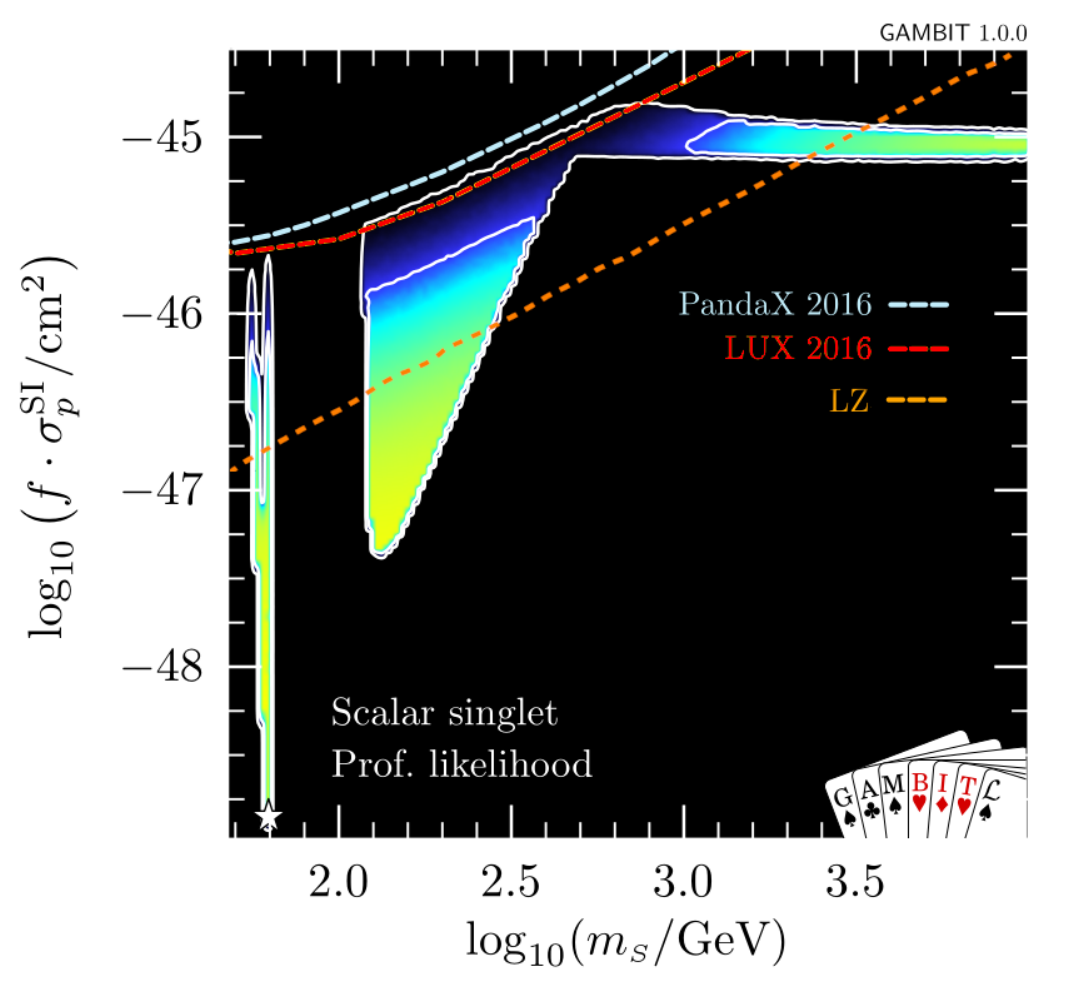}
\caption{Profile likelihood of the scalar singlet DM model in the mass $m_S$ vs coupling $\lambda_{hS}$ plane, for the low mass range (left) and the high mass range (centre). The right-hand panel shows the spin-independent cross section, with overlaid exclusions from PandaX 2016, LUX 2016 and LZ 2022.}
\label{fig:scalarsinglet}
\end{figure}

Figure \ref{fig:scalarsinglet} show the profile likelihood results for the scalar singlet model where the star marks the best fit point and the white contours at 1$\sigma$ and 2$\sigma$ confidence intervals. The shape of the contours is mostly driven by the upper limit on the relic density, for small couplings, and by direct detection constraints for large couplings. In fact, from the diret detection constraints, PandaX 2016 and LUX 2016 are the most constraining, as can be seen in the right-hand panel of Figure \ref{fig:scalarsinglet}. We also overlay the expected exclusion from the results of the LZ 2022 experiment, which were not included in the scan at the time; it would exclude roughly half of the remaining parameter space for high masses.

\begin{figure}[ht]
\includegraphics[width=0.33\textwidth]{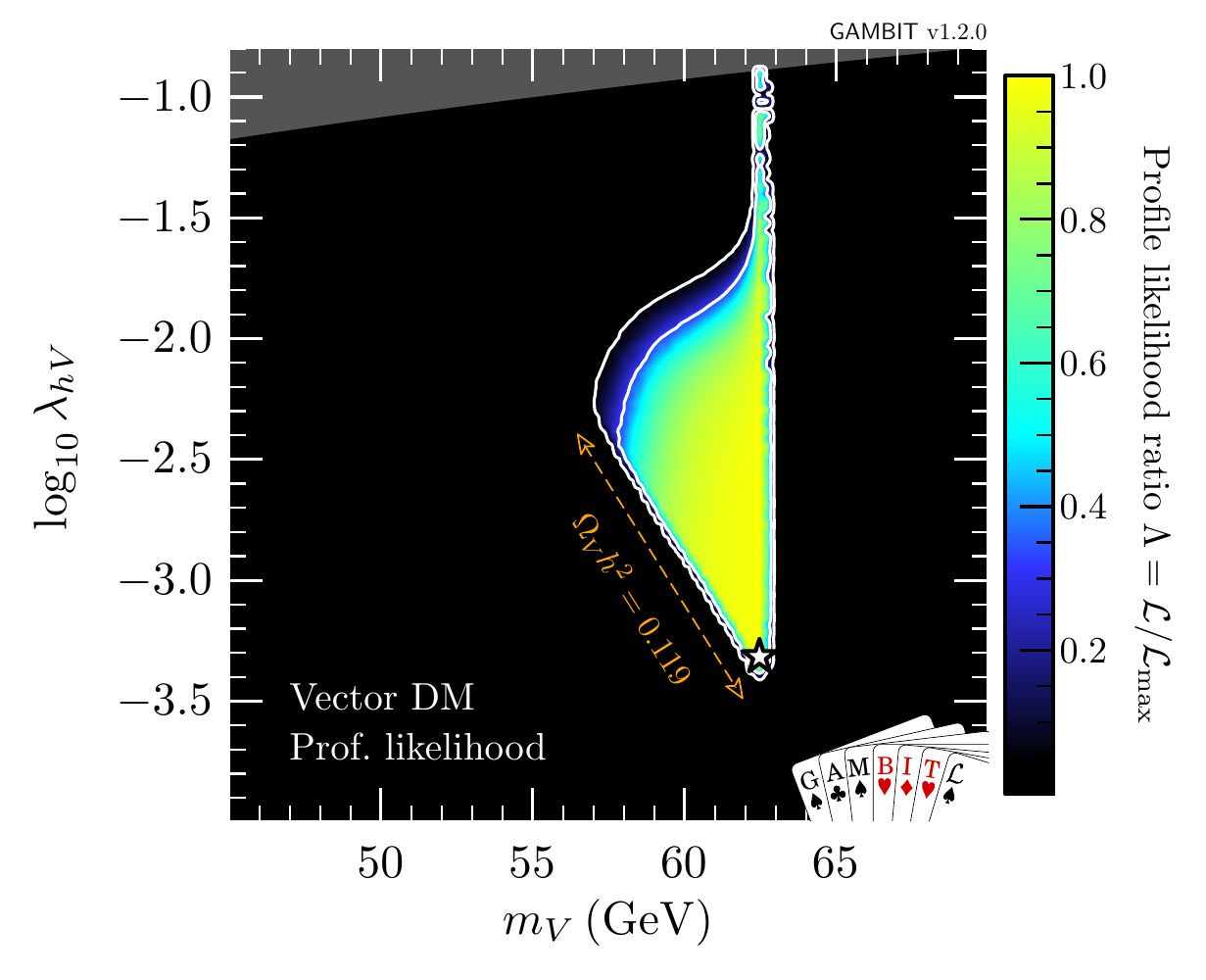}
\includegraphics[width=0.33\textwidth]{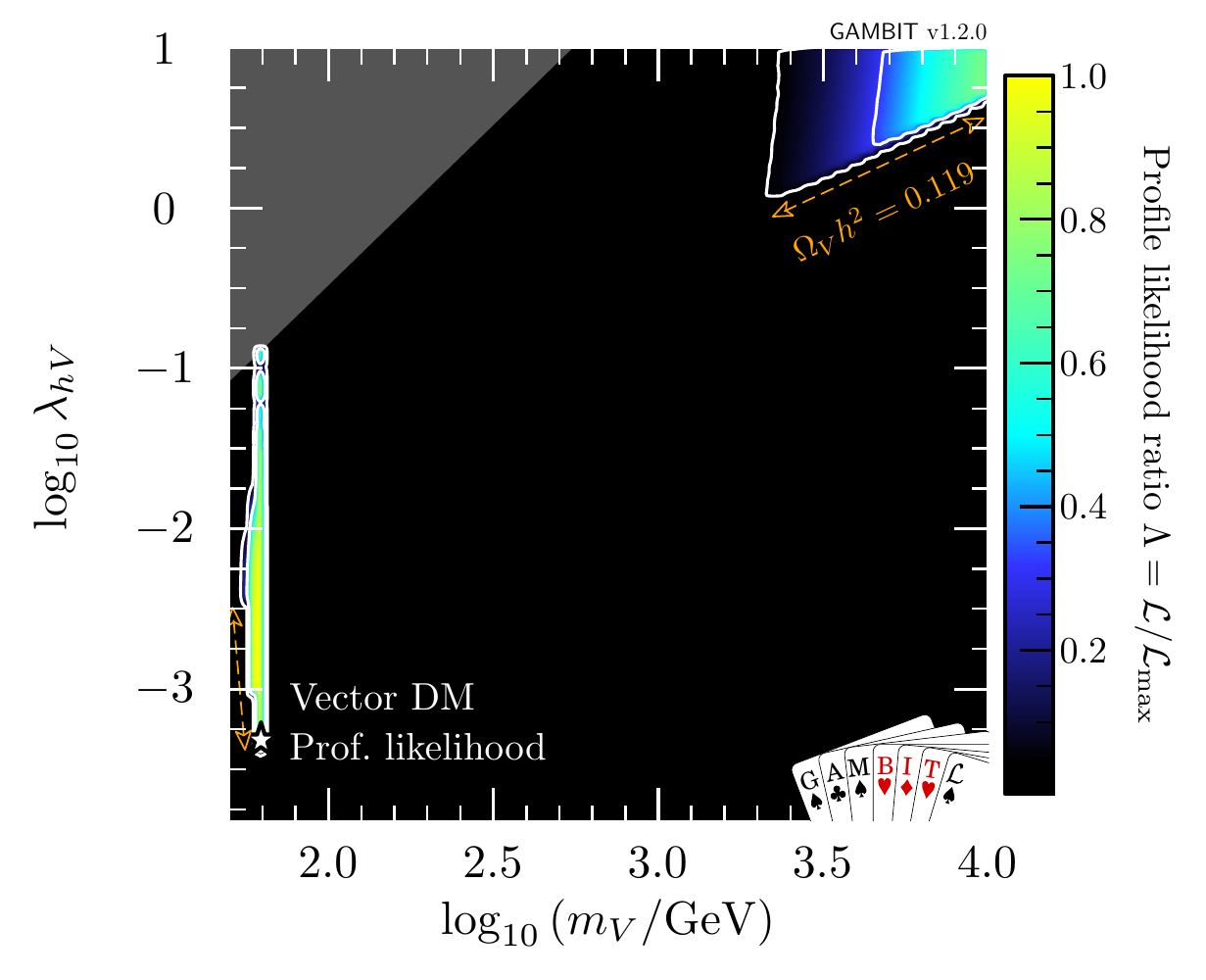}
\includegraphics[width=0.33\textwidth]{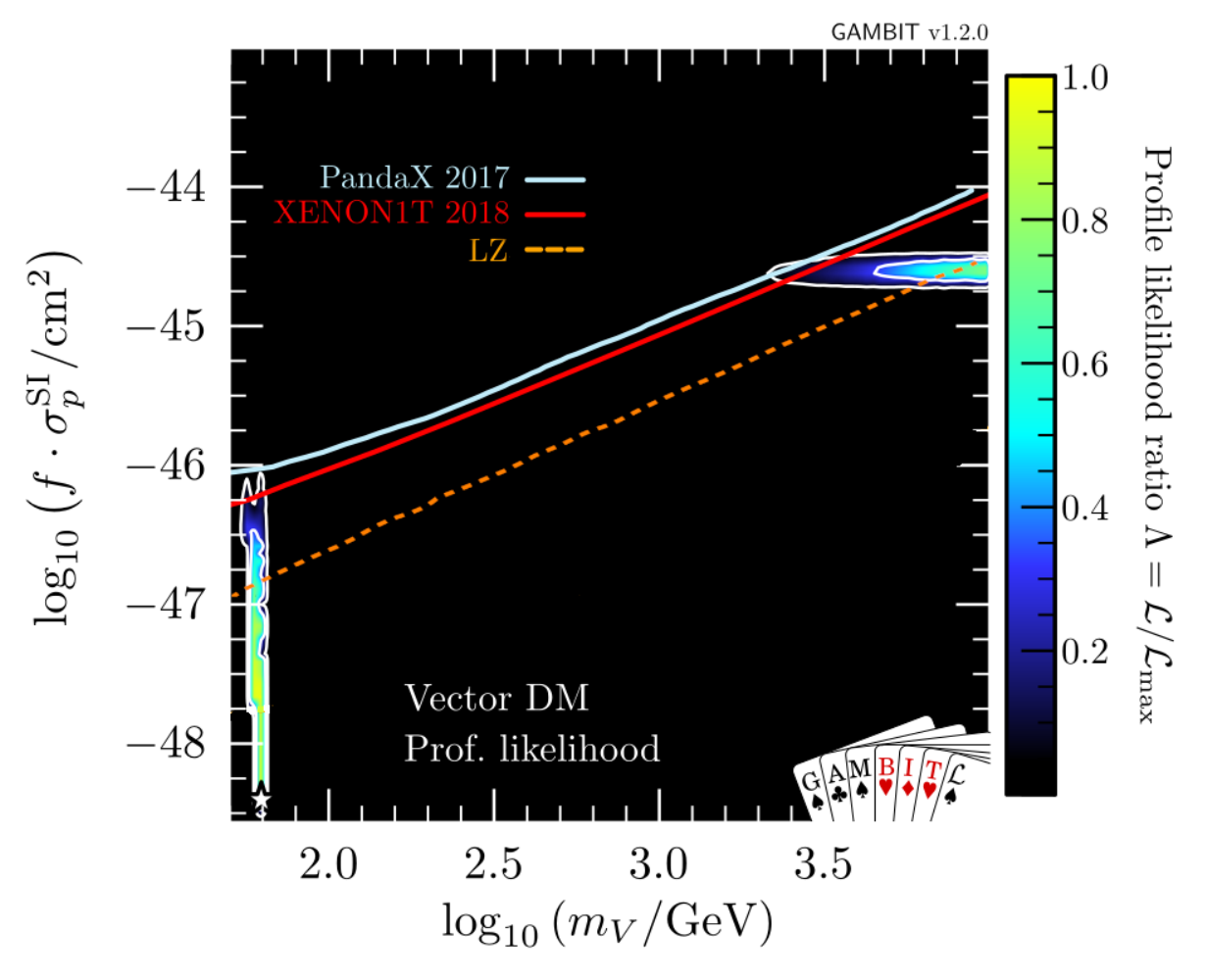}
\caption{Profile likelihood of the vector singlet DM model in the mass $m_X$ vs coupling $\lambda_{hV}$ plane, for the low mass range (left) and the high mass range (centre). The right-hand panel shows the spin-independent cross section, with overlaid exclusions from PandaX 2016, LUX 2016 and LZ 2022.}
\label{fig:vectorsinglet}
\end{figure}

Figure \ref{fig:vectorsinglet} shows the equivalent profile likelihood plots for the vector DM model, for low mass (left) and high mass (centre). As it is expected the results are fairly similar to the scalar singlet DM, with two notable differences. First there is a large grey shaded region, which is excluded by perturbative unitarity. Second, the contours are much smaller, particularly for the high mass region, which is mostly due to the stronger direct detection constraints from XENON1T and PandaX 2017. This is evident on the right-hand plot where the constraints from the newer direct detection experiments are overlaid. We also show the predicted exclusions from LZ, which clearly would exclude most of the high mass region.

\begin{figure}[ht]
\centering
\includegraphics[width=0.4\textwidth]{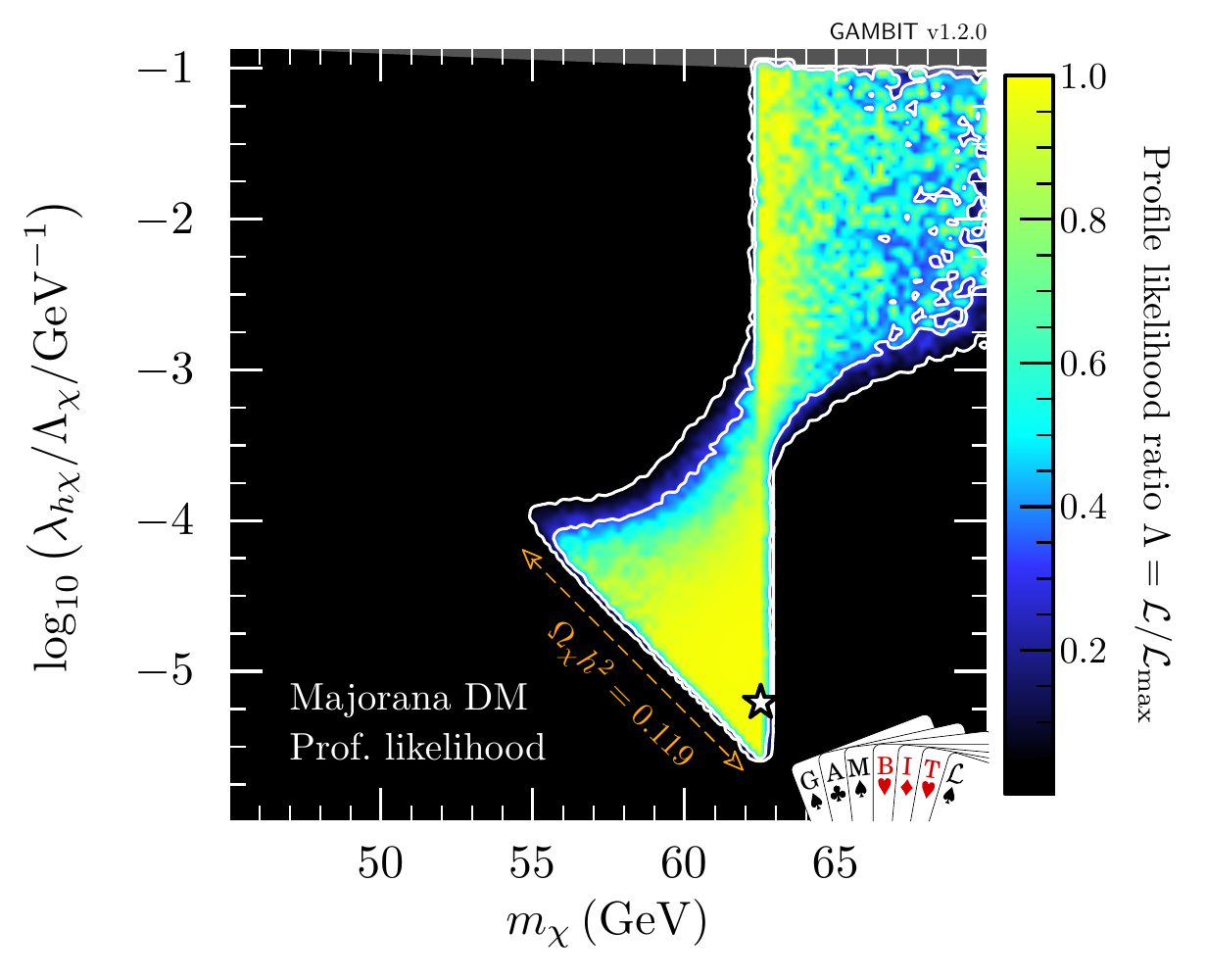}
\includegraphics[width=0.4\textwidth]{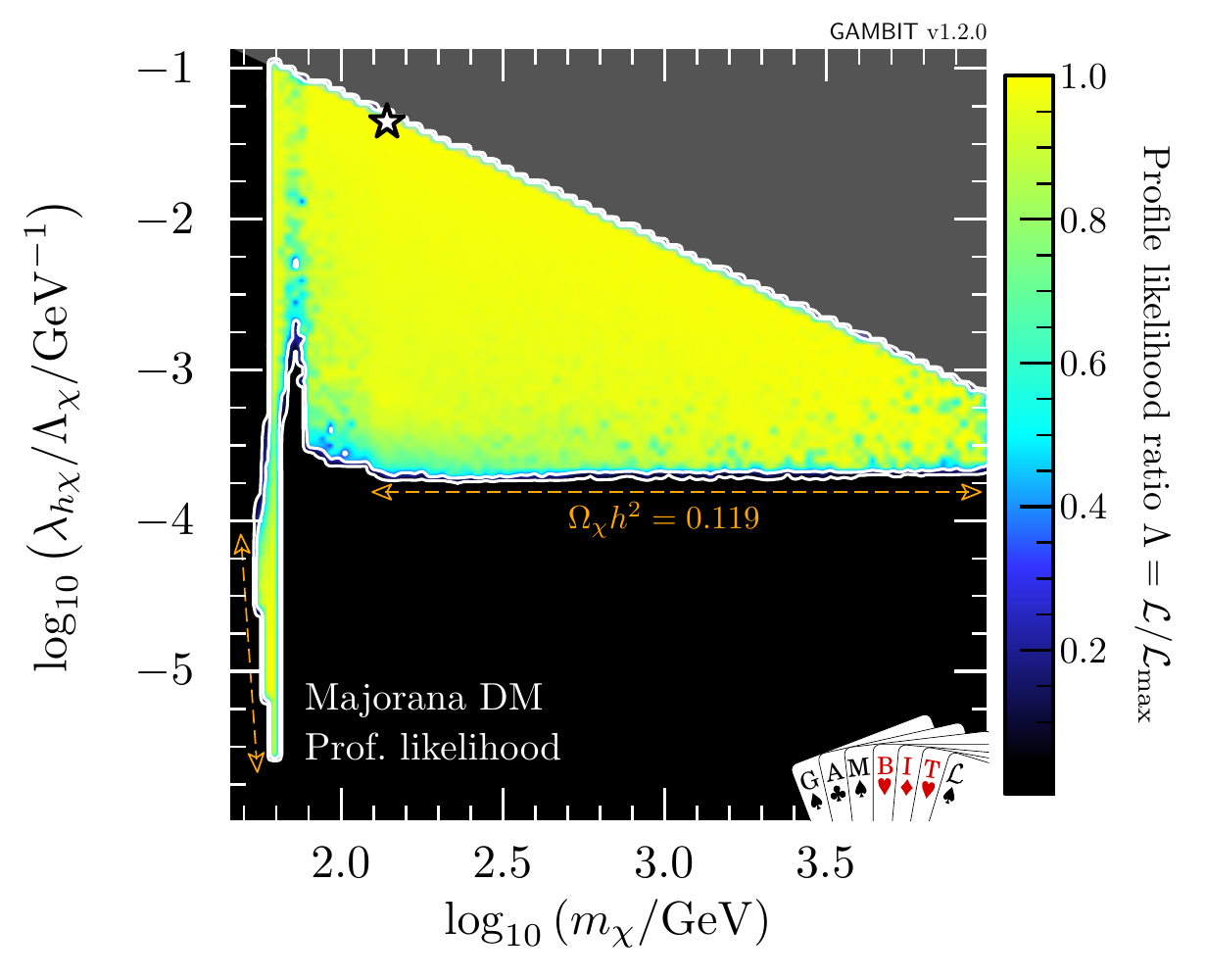}
\includegraphics[width=0.4\textwidth]{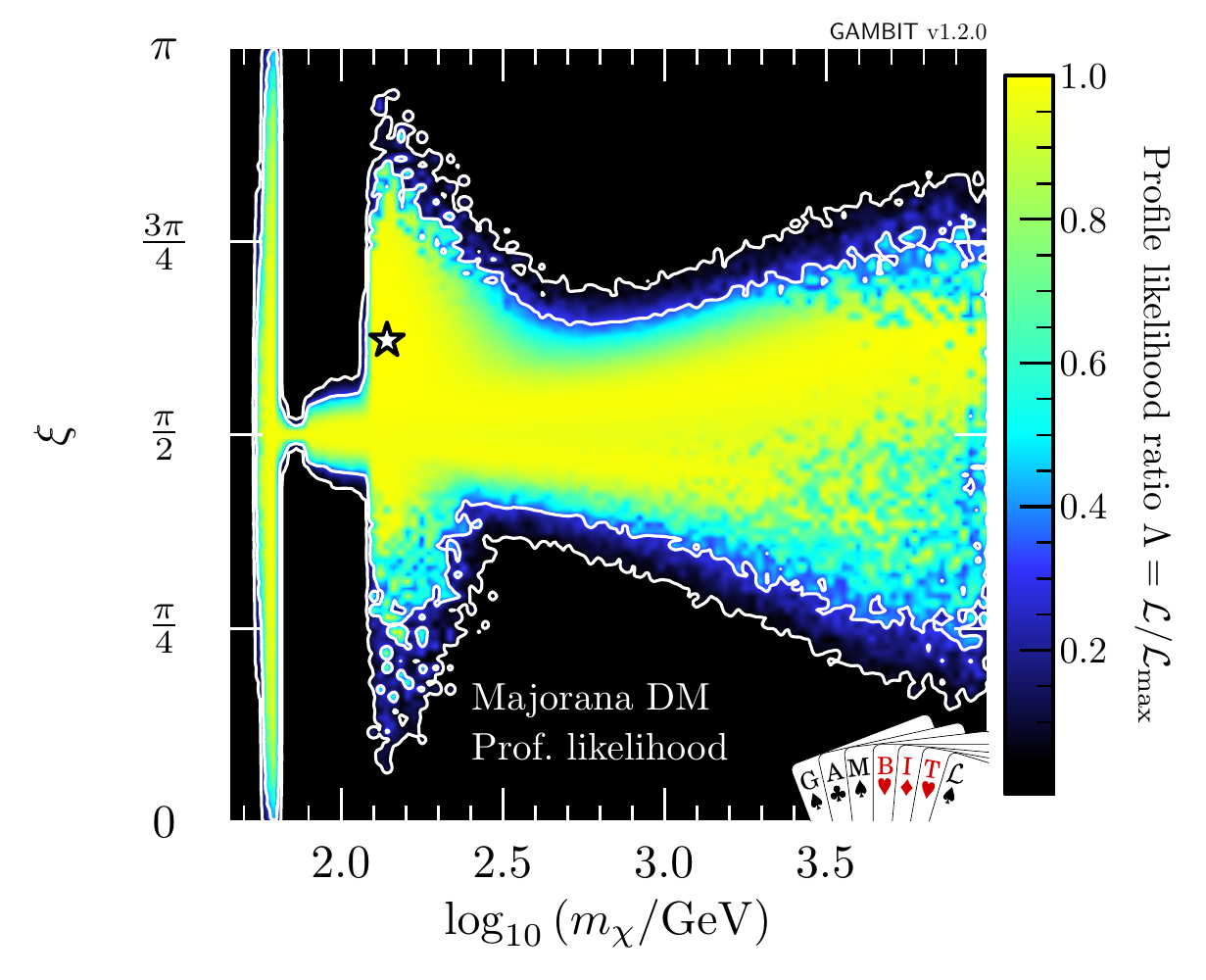}
\includegraphics[width=0.4\textwidth]{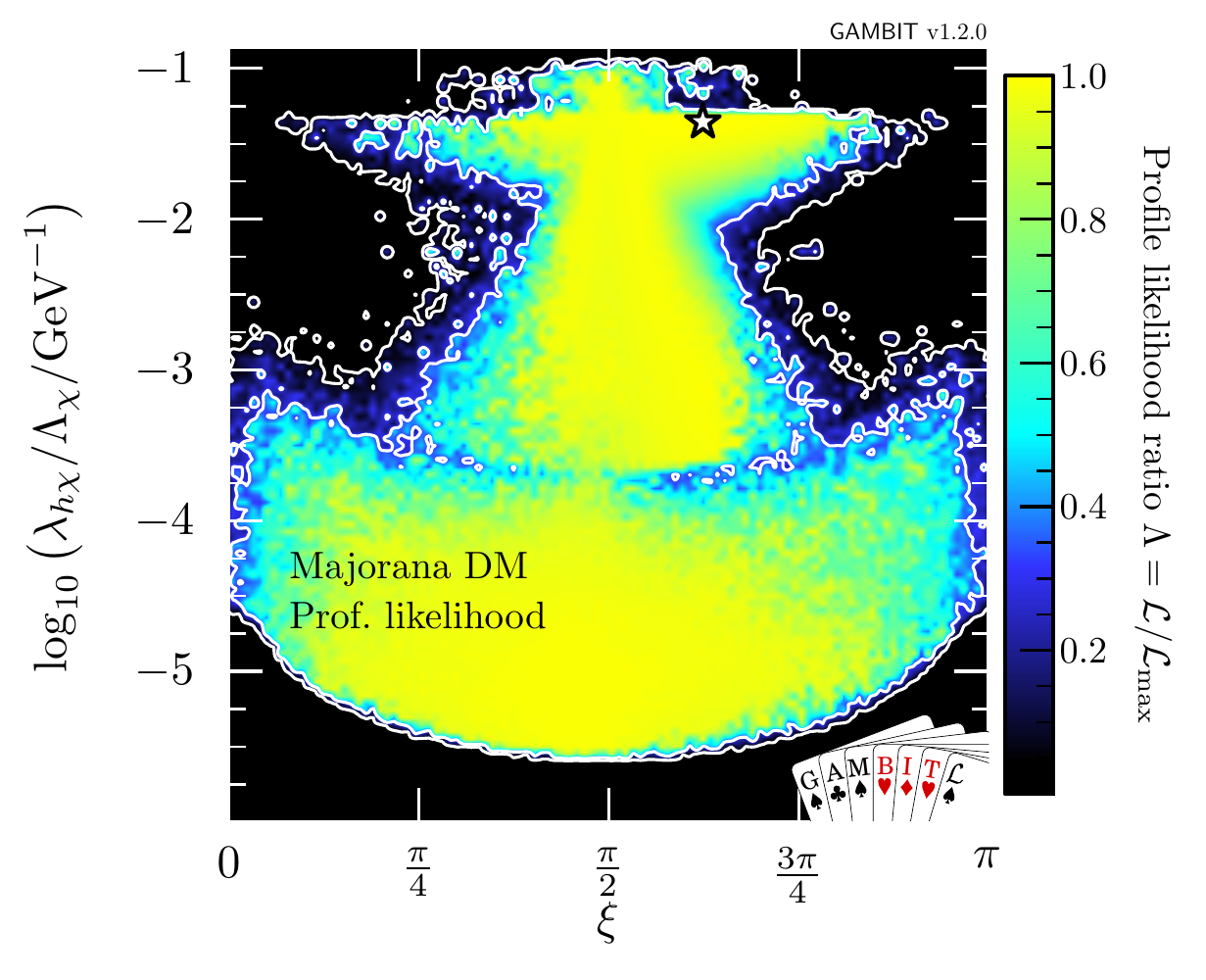}
\caption{Profile likelihood of the Majorana fermion singlet DM model for combinations of the mass $m_\chi$, coupling $\lambda_{h\chi}/\Lambda_\chi$ and CP phase $\xi$, for the low mass range (top left) and the high mass range (top right).}
\label{fig:fermionsinglet1}
\end{figure}

Lastly, Figure \ref{fig:fermionsinglet1} shows the profile likelihood results for the Majorana fermion singlet model. We do not show here the results for the Dirac fermion model, as they are effectively identical and do not add further information. The most notable difference with respect to the bosonic DM models is that the low and high mass regions are connected. This is due to the CP phase $\xi$\footnote{The CP phase $\xi$ is defined as $\xi = \theta + \alpha$ where $\alpha$ a field rotation to eliminate unwanted pseudoscalar bilinears after EWSB.}, as can be seen in the bottom left-hand panel, where the connection between the low mass and high mass regions occurs mostly for maximal CP violation $\xi = \pi/2$. This is primarly due to the suppression of the direct detection constraints, as a consequence of the pseudoscalar nature of the interaction for $\xi \sim \pi/2$. As opposed to the scalar and vector models, the inclusion of the recent LZ constraint would not exclude large regions of the parameter space, as the direct detection constraints are suppressed.

\section{DM Effective Field Theory}
\label{sec:DMEFT}

An alternative approach to specific models with explicit couplings between the DM particles and the SM, as the Higgs-portal models above, one can parametrise the interactions with an EFT approach. In this paradigm we consider a Dirac fermion DM candidate $\chi$ with the following Lagrangian
\begin{equation}
 \mathcal{L} = \mathcal{L}_{\rm SM} + \mathcal{L}_{\rm int} + \bar\chi(i\slashed\partial - m_{\chi})\chi,
\end{equation}
where $\mathcal{L}_{\rm SM}$ includes the SM-only Lagrangian terms, and $\mathcal{L}_{\rm int}$ describes the interactions between the DM particle and the SM, containing a series of high dimensional operators $\Q{a}{d}$, of the following form
\begin{equation}
 \mathcal{L}_{\rm int} = \sum_{a,d}\frac{\C{a}{d}}{\La^{d-4}}\Q{a}{d},
\end{equation}
where $\C{a}{d}$ are the Wilson coefficents (WCs) for each operator and $\La$ a common scale of new physics.

For simplicity we assume that there are no interactions between DM particles and leptons This is motivated because direct detection signals (often the most relevant) are due to the scattering of DM particles with nuclei, which corresponds to interactions with quarks and gluons. Therefore, the 6 and 7-dimensional operators that we consider in this study are
\begin{equation}
 \begin{array}{ll}
   & \Q{1}{7} = \frac{\alpha_s}{12\pi}(\overline\chi \chi)G^{a\mu\nu}G^a_{\mu\nu}\,,\\
   & \Q{2}{7} = \frac{\alpha_s}{12\pi}(\overline\chi i\gamma_5 \chi)G^{a\mu\nu}G^a_{\mu\nu}\,,\\
   & \Q{3}{7} = \frac{\alpha_s}{8\pi}(\overline\chi \chi)G^{a\mu\nu}\widetilde{G}^a_{\mu\nu}\,, \\
   \Q{1,q}{6} = (\overline\chi \gamma_\mu \chi)(\overline{q} \gamma^\mu q)\,, & \Q{4}{7} = \frac{\alpha_s}{8\pi}(\overline\chi i\gamma_5 \chi)G^{a\mu\nu}\widetilde{G}^a_{\mu\nu}\,, \\
   \Q{2,q}{6} = (\overline\chi \gamma_\mu \gamma_5 \chi)(\overline{q} \gamma^\mu q)\,,  &    \Q{5,q}{7} = m_q(\overline\chi \chi)( \overline{q} q )\,, \\
   \Q{3,q}{6} = (\overline\chi \gamma_\mu \chi)(\overline{q} \gamma^\mu \gamma_5 q)\,, & \Q{6,q}{7} = m_q(\overline\chi i\gamma_5 \chi)( \overline{q} q )\,, \\
   \Q{4,q}{6} = (\overline\chi \gamma_\mu \gamma_5 \chi)(\overline{q} \gamma^\mu \gamma_5 q)\,. &    \Q{7,q}{7} = m_q(\overline\chi \chi)( \overline{q} i\gamma_5 q )\,, \\
   &    \Q{8,q}{7} = m_q(\overline\chi i\gamma_5 \chi)( \overline{q} i\gamma_5 q )\,, \\
    &    \Q{9,q}{7} = m_q(\overline\chi \sigma^{\mu\nu} \chi)( \overline{q} \sigma_{\mu\nu} q )\,, \\
    &    \Q{10,q}{7} = m_q(\overline\chi i\sigma^{\mu\nu}\gamma_5 \chi)( \overline{q} \sigma_{\mu\nu} q )\,.\\

  \label{operators}
  \end{array}
 \end{equation}
The WCs associated with each operator are defined at the input scale $\Lambda$, but they are required at various other scales for the purpose of computing constraints. In particular, direct detection constraints require the WCs at $\mu = 2$ GeV, and we compute this running of the WCs using \textsf{DirectDM}~\cite{Bishara:2017nnn}. In addition, for $\La > m_t$, WCs get a contribution from mixing of the form
\begin{align}
 \C{1,2}{5} &= 4 \frac{m_t^2}{\La^2}\log\frac{\La^2}{m_t^2}\C{9,10}{7}, \notag \\
 \Delta \C{i}{7} &= -\C{i+4,q}{7} (i=1,2), \quad\quad \Delta \C{i}{7} = \C{i+4,1}{7} (i=3,4).
\end{align}

\begin{figure}{ht}
\centering
 \includegraphics[width=0.6\textwidth]{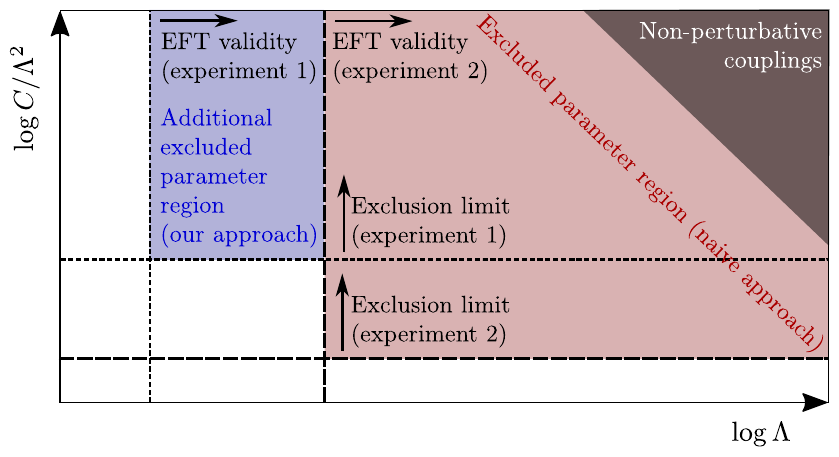}
 \caption{Illustrative coverage of the EFT parameter space, highlighting our approach (blue) over the naive approach (red).}
 \label{fig:eftvalidity}
\end{figure}

The issue of the EFT validity is critical in the study of of these models. In our study we have decided to leave $\La$ a free parameter, instead of combining it with the WCs as $C/\La$ as the naive approach. This allows us to cover more of the potential parameter space of the model, as schematically shown in Figure \ref{fig:eftvalidity}. By taking $\La$ as a free parameter we are also able to set constraints on the validity of the EFT. Direct detection constraints require a lower limit of $\La > 2$ GeV on the scale of new physics. Furthermore, in order to have sensible annihilations we also require that $\La > 2m_\chi$. Lastly, since we include collider constraints on the model, we impose EFT validity as a cut-off on the missing energy spectrum. We do not change the spectrum for $\slashed{E}_T < \La$, but when $\slashed{E}_T > \La$ we take two approaches, a hard cut-off, where we set $d\sigma/d\slashed E_T = 0$, and a smooth cut-off where we make the replacement $\frac{d\sigma}{d\slashed E_T} \to \frac{d\sigma}{d\slashed E_T}\left(\frac{\slashed E_T}{\La}\right)^{-a}$, where we keep the slope of the smooth cut-off parameter $a$ as a free parameter in our study.

Regarding the constraints we include in this study, they are very similar to those for the Higgs-portal models above, so here we only describe the differences. For direct detection, we perform a matching from the relativistic operators in eq.~\eqref{operators} to the non-relativistic operators required for direct detection with \textsf{DirectDM} and compute the likelihood contributions with \textsf{DDCalc}. This study includes also the results from PICO-60 2019~\cite{PICO:2019vsc}. To compute the annihilation cross section for this model we have autogenerated the required code with \textsf{GUM}~\cite{Bloor:2021gtp}, computed the cross sections with \textsf{CalcHEP}~\cite{Belyaev:2012qa} and the relic abundance with \textsf{DarkSUSY}, comparing it to the 2018 measurement of the relic density by the Planck satellite, $\Omega_{DM}h^2 \leq 0.120 \pm 0.001$~\cite{Planck:2018vyg}. Indirect detection constraints are computed as in the Higgs-portal case above. Annihilations of DM in the early Universe may inject energy into the primordial plasma and affect the CMB, which can be parametrised with a constraint on the energy deposition efficiency $f_{\rm eff}$, computed with \textsf{DarkAges}~\cite{Stocker:2018avm,Renk:2020hbs}. Lastly we include constraints from monojet searches at ATLAS~\cite{ATLAS:2021kxv} and CMS~\cite{CMS:2017zts}. We have generated interpolated tables of cross sections and efficiency-acceptance based on simulations using \textsf{MadGraph\_aMC@NLO}~\cite{Alwall:2011uj} and \textsf{Pythia}~\cite{Sjostrand:2007gs}. As there are some minor bin-wise excesses on these searches, we analyse our results with either a capped LHC likelihood (exclusion only) or uncapped (allowed to freely fit upwards fluctuations).

\begin{figure}[ht]
\centering
\includegraphics[width=0.35\textwidth]{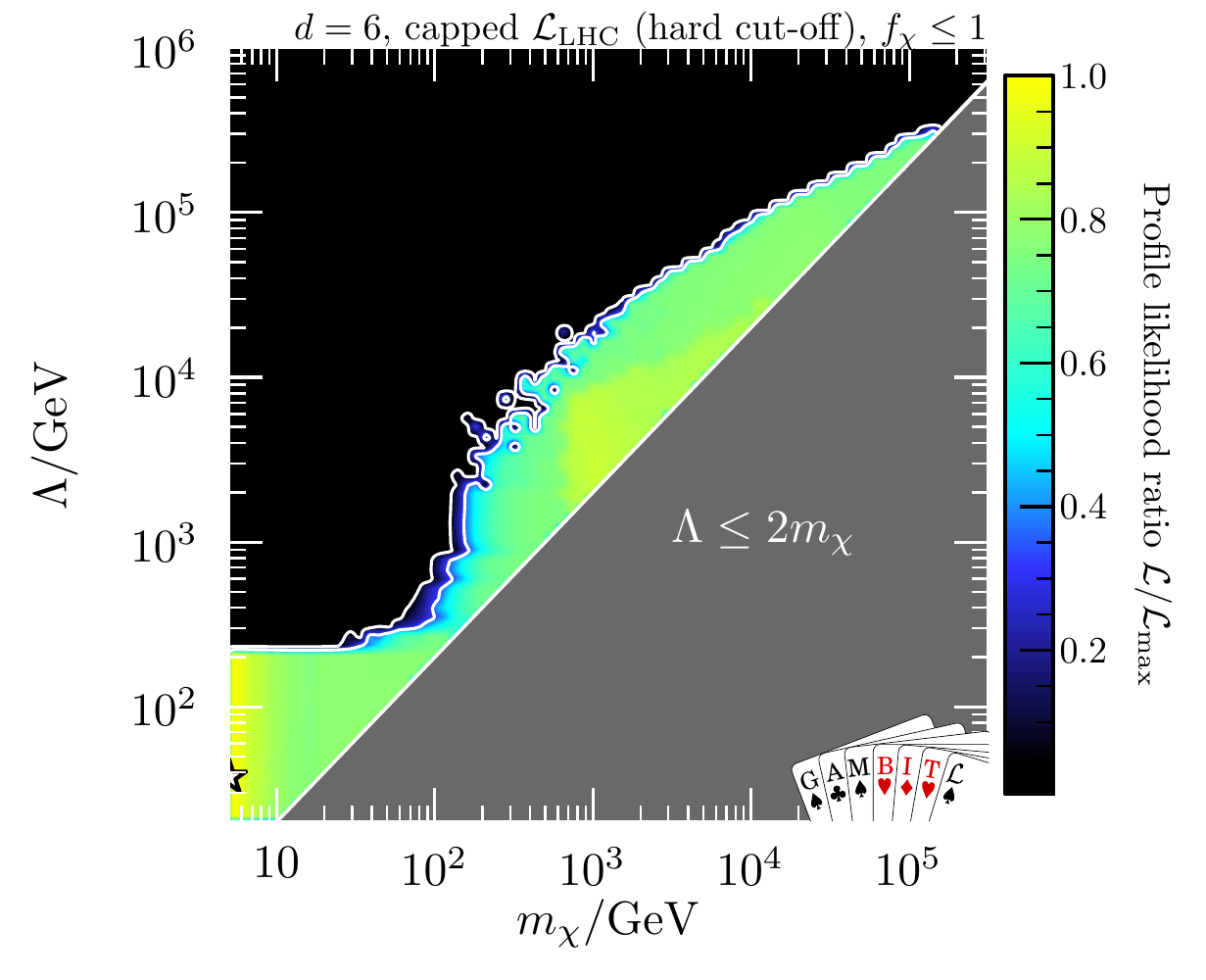}
\includegraphics[width=0.35\textwidth]{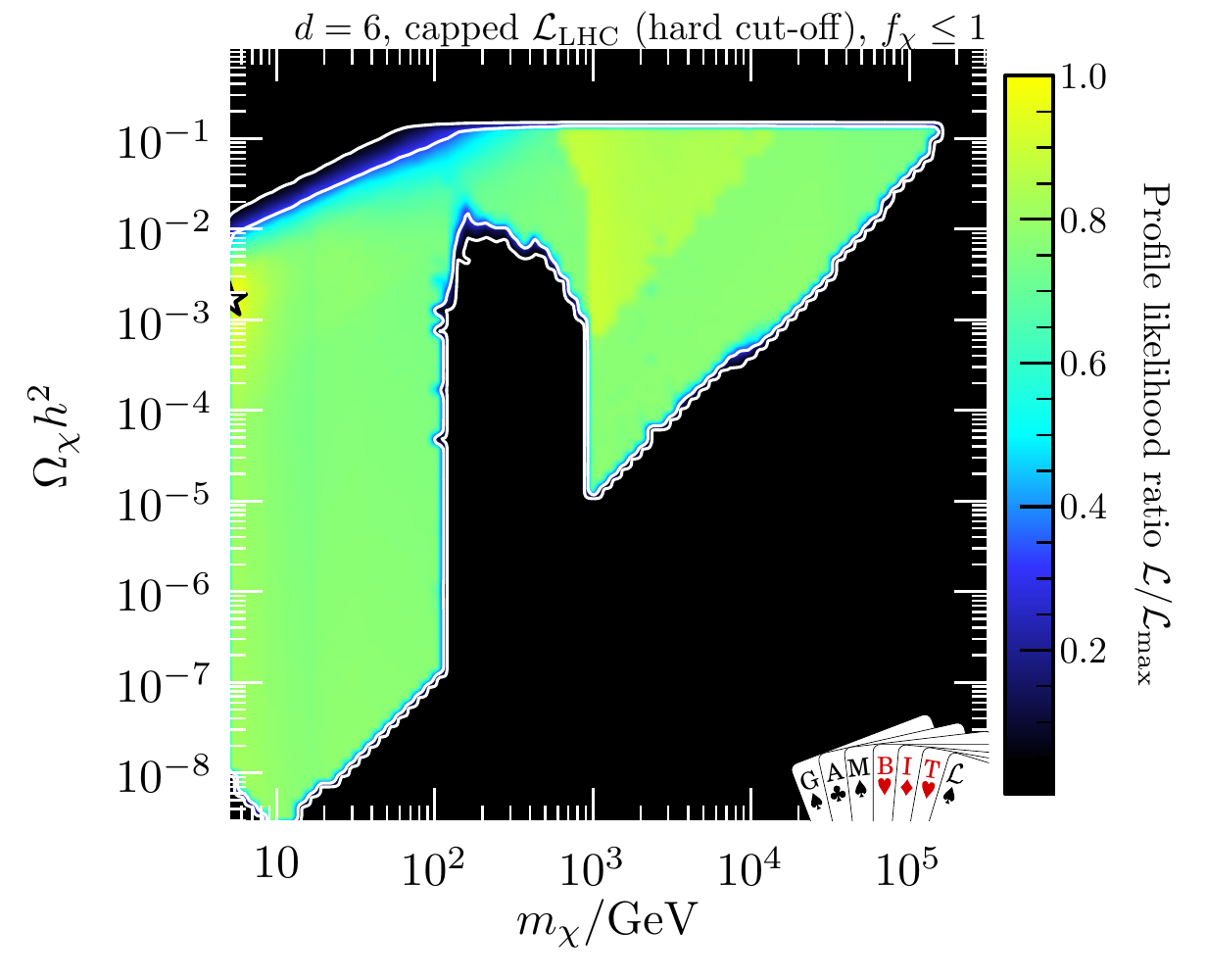}\\
\includegraphics[width=0.35\textwidth]{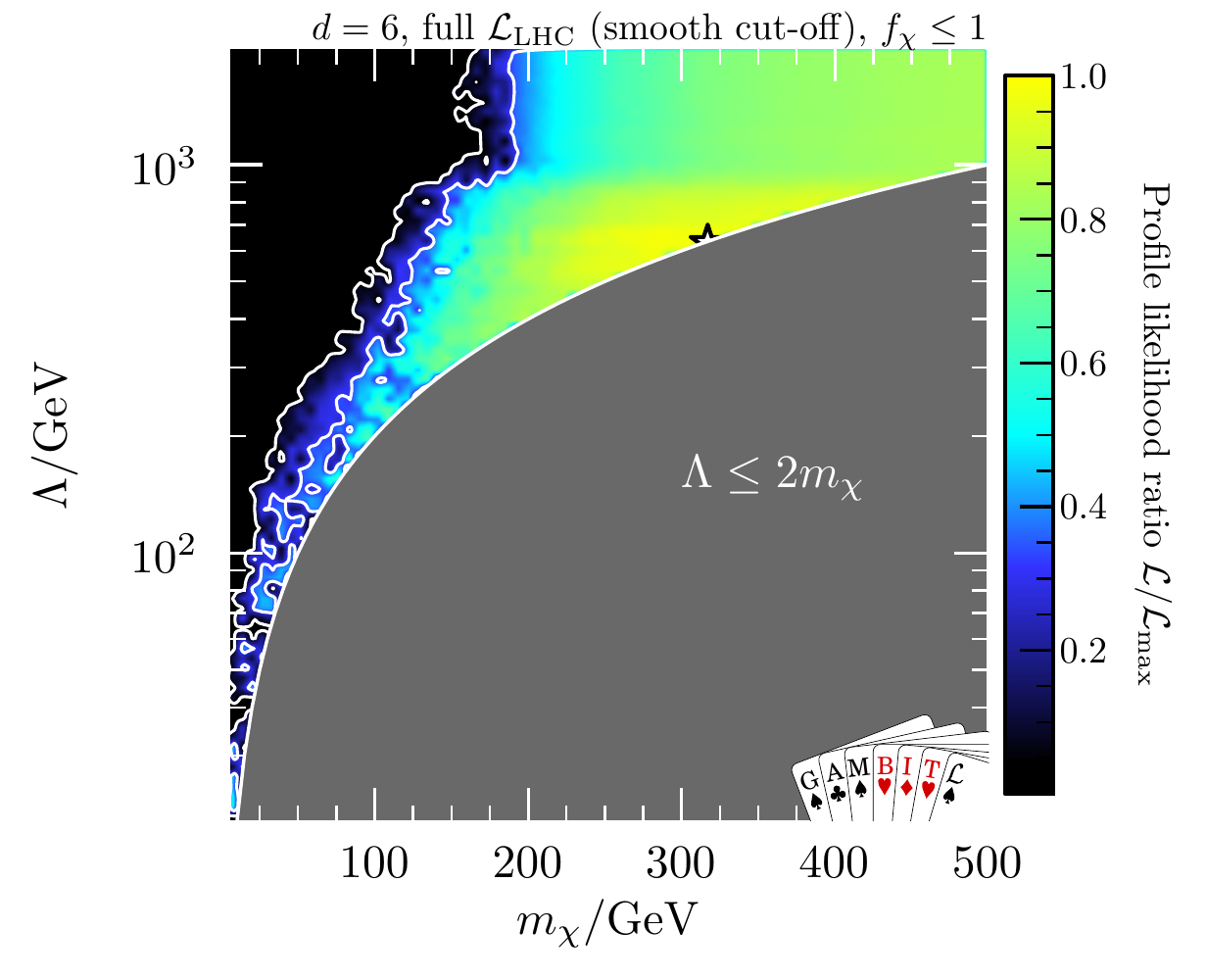}
\includegraphics[width=0.35\textwidth]{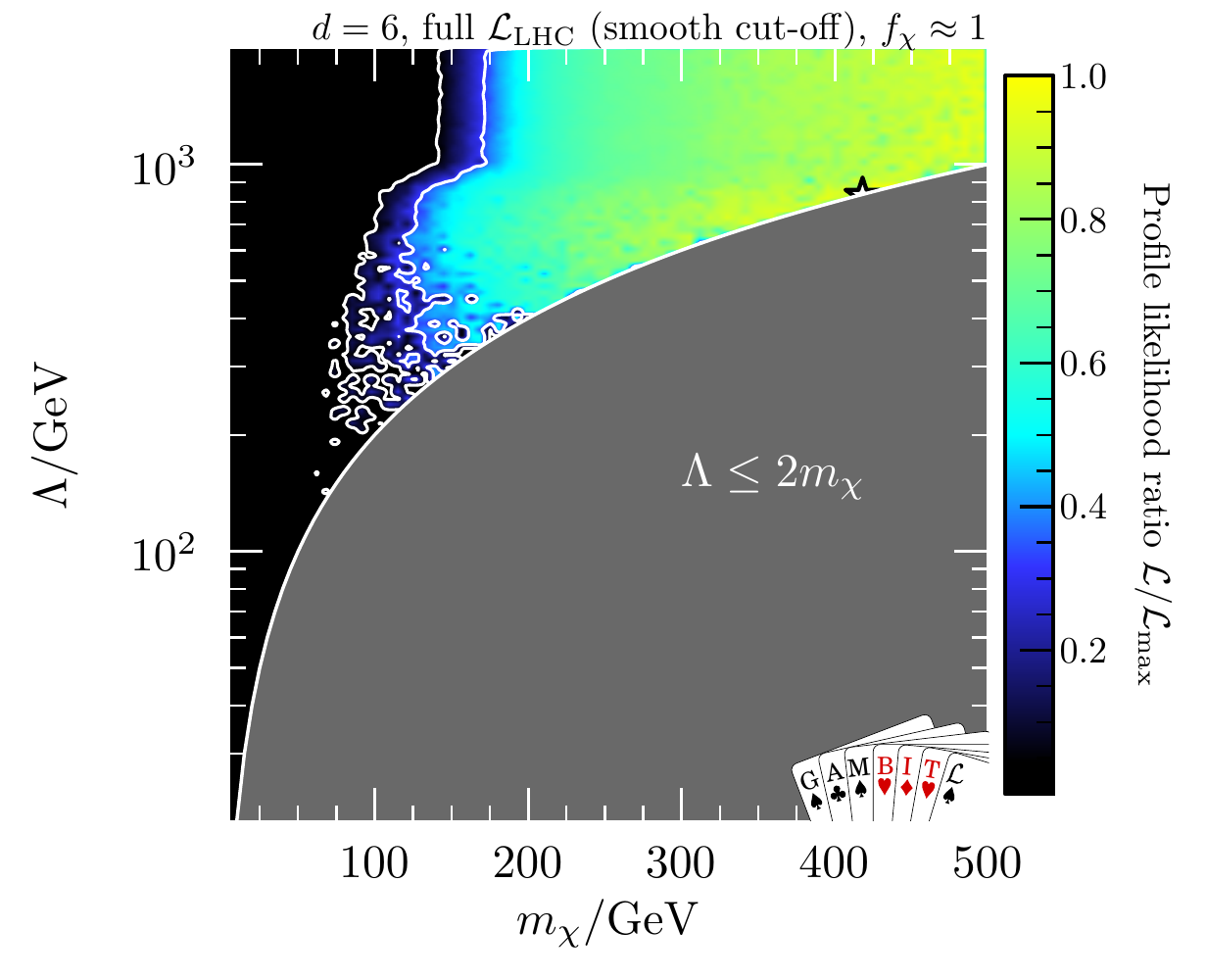}
\caption{Profile likelihood in the $m_\chi - \La$ plane (top-left and bottom) and the $m_\chi-\Omega_\chi h^2$ plane (top-right). In the top row we show the results when the LHC likelihood is capped, and the bottom row when it is allowed to fit the excesses, with underabundant (left) or exact (right) relic density.}
\label{fig:DMEFT}
\end{figure}

We show in Figure \ref{fig:DMEFT} the results for a scan with all dimension 6 operators from eq.~\eqref{operators}. For the results with dimension 7 operators we refer the reader to the original paper~\cite{GAMBIT:2021rlp}. One of the most interesting results that can be extracted from these plots (top-left) is that there is an upper limit on the scale of new physics of around $\La < 300$ TeV, due to the the requirement of thermal freeze-out. The top-right panel shows that it is not possible to saturate the relic density for masses below $m_\chi \lesssim 100$ GeV, from a combination of direct and indirect detection contraints. The bottom-left panel shows that there is a slight preference for masses around $m_\chi \approx 300$ GeV when the model can fit the small excesses seen in ATLAS and CMS searches. This preferred mass shifts to much higer values (around $m_\chi \approx 400$ GeV), when DM saturates the observed relic abundance.

\section{Conclusions}
\label{sec:conclusions}

In this conference paper I showed results from a global fit of various Higgs-portal DM  models and a DM EFT model. The results showed that bosonic Higgs-portal models are in strong tension with experiments, where mostly only the low mass region around the Higgs resonance survives, due to recent direct detection results. Fermionic DM (Majorana or Dirac) does not suffer the same fate, as direct detection signals are suppressed when the CP violation phase is maximal. The study of the DM EFT model showed that there is still a lot of parameter space available for models that can be matched onto an EFT. Most notable, if WIMPs are the right candidate for DM, there is maximum scale of new physics where they should appear.

\pagebreak

\end{document}